# FRACTAL PROPERTIES IN ECONOMICS

HIDEKI TAKAYASU

*Sony Computer Science Laboratories Inc.,3-14-13, Higashigotanda, Shinagawa-ku,*
*Tokyo, 141-0022, Japan*
*E-mail:takayasu@csl.sony.co.jp*

MISAKO TAKAYASU

*Faculty of Science and Technology, Keio University,*
*890-12 Kashimada, Saiwai-ku, Kawasaki 211-0958, Japan*
*E-mail:misako@future.st.keio.ac.jp*

MITSUHIRO P. OKAZAKI

*Department of Industrial Engineering, Musashi Institute of Technology,*
*1-28-1, Tamazutsumi, Setagaya-ku, Tokyo, 158-8557, Japan*
*E-mail:f6586719@hi-ho.ne.jp*

KOUHEI MARUMO

*Institute for monetary and economic studies, Bank of Japan*
*2-1-1 Hongoku-cho Nihonbashi, chuo-ku, Tokyo, 103-8660, Japan*
*E-mail:kouhei.marumo@boj.or.jp*

TOKIKO SHIMIZU

*Financial markets department, Bank of Japan*
*2-1-1 Hongoku-cho Nihonbashi, chuo-ku, Tokyo, 103-8660, Japan*
*E-mail:tokiko.shimizu@boj.or.jp*

Scaling properties in financial fluctuations are reviewed from the standpoint of statistical physics. We firstly show theoretically that the balance of demand and supply enhances fluctuations due to the underlying phase transition mechanism. By analyzing tick data of yen-dollar exchange rates we confirm two fractal properties: 1 The distribution of rate change in a fixed ticks is approximated by a symmetric stretched exponential function for a wide range of time intervals; 2 the interval time distribution of trades nearly follows a power law. Empirical fractal properties in companies' financial data, such as distributions and fluctuations in assets and incomes are discussed with a simple model. The importance of methods and theories for phase transitions is discussed.

## 1  Introduction

The concept of fractals has been spread over all fields of sciences and technology, however, it seems less known that this concept was born in the field of economics when Mandelbrot was investigating the price changes in an open market around 1963 [1]. He found empirically that a chart of market price changes of cotton price looks similar to another chart with different time resolution. A general concept flashed across his mind that such scale invariance could be the clue to characterize many complex phenomena around us. After about 20 years the concept he created, the fractals, became a key word in a wide research activity. Now, more than 35 years have passed since his original discovery, economics becomes one of the hottest topics in the study of fractals.

A new field of science, econophysics, was established recently in 1997 [2, 3]. This is a study of economic phenomena based on the methods and approaches of physics. Among many topics in econophysics there are three topics that are closely related to the study of fractals. They are price changes in open market, the distribution of income of companies, and the scaling relation of company's size fluctuations. In this paper we briefly review these topics and discuss about the validity of the known empirical laws introducing simple mathematical models for better understanding.

In the following chapter we discuss about price fluctuations which naturally involve critical fluctuations near the equilibrium point of demand and supply. In the 3rd chapter we show results of tick data analysis of foreign exchange rates. We show theoretically that the fat tails of the rate change distribution can be modeled by a very simple physical equation of Langevin equation with random coefficient. In the 4th chapter we review the empirical laws on companies' macroscopic financial data of annual income and assets. We introduce a very simple model of company growth that reproduces the Zipf's law in income distribution, also we discuss a nonlinear relation between income and assets. The 5th chapter is devoted for general discussion and concluding remarks on the importance of statistical physics for the data analysis in economics.

## 2  Demand and supply; a phase transition view

It had been a common sense in economics for a long time that demand and supply balances automatically, however, it becomes evident that in reality such balances are hardly be realized for most of popular commodities in our daily life [4, 5]. The important point is that demand is essentially a stochastic variable because human action can never be predicted perfectly, hence the balance of demand and supply should also be viewed in a probabilistic way. If demand and supply are balanced on average the probability of finding an arbitrarily chosen commodity on the shelves of a store should be 1/2, namely about half of the shelves should be empty. Contrary to this theoretical estimation shelves in any department store or supermarket is nearly always full of commodities. This clearly demonstrates that supply is much in excess in such stores. Excess supply generally holds for most of commodities especially foods in economically advanced countries. For example, it is reported in a newspaper that about a quarter of foods produced in the USA are disposed as garbage. If food supply were equal to demand on average then it implies that about half people were in starvation!

In general the stochastic properties of demand and supply can be well characterized by a phase transition view which is consisted of two phases; the excess-demand and excess-supply phases. It has been mathematical proven that for commodities of which production costs are smaller than a certain portion of the selling prices, an excess-supply strategy becomes the best strategy in the situation that demand fluctuates randomly and unsold commodities are to be disposed [6]. Obviously department stores follow this excess-supply strategy. The other strategy of keeping excess-demand can also be a rational solution when the production cost and the selling price are close. The critical point at which the averaged demand and supply are equal realizes only in a very special case.

It is a general property of a phase transition system that fluctuations are largest at the phase transition point, and this property also holds in this demand-supply system. In the case of markets of ordinary commodities, consumers and providers are independent and the averaged supply and demand are generally not equal. The resulting price fluctuations are generally slow and small in such market because the system is out of the critical point. On the contrary in an open market of stocks or foreign exchanges, market is governed by speculative dealers who frequently change their positions between buyers and sellers. It is shown that such speculative actions make demand and supply balance automatically on average by changing the market price [5]. As the system is always at the critical point the resulting price fluctuations are generally quick and large involving fractal properties.

## 3    Price changes in open markets

The physical mechanism of market price fluctuations was firstly studied by one of the authors (H. T.) et al in '92 by introducing a numerical market model[7]. The model consists of speculative dealers who transact with others simply following the basic rule "buy at a lower price and sell at a higher price". These two threshold prices are determined at each time step by each dealer taking into account the information of past market price changes. It is shown that even a smallest limit case of 3 dealers can show chaotic behaviors, implying that the transaction's nonlinear effect is very strong. The resulting market prices naturally include sudden falls or increases when the parameter controlling the dealers' averaged response to latest price change is large enough.

In '95 a famous paper by Mantegna and Stanley appeared in Nature in which price changes in a stock market are reported to follow a power law distribution with truncation at higher values [8]. The exponent of the power law is estimated as 1.4 in the form of stable distribution's characteristic exponent. However, this power law has not been accepted widely and the precise functional form of the distribution of open market price changes is still under intensive debate.

Here, we show our original results of analysis on tick data of yen-dollar exchange rates for about six months. Tick data are the highest resolution transaction data of foreign exchange markets, which are announced to the dealers electronically. Transactions of foreign exchanges are done by telephone calls, so the market is active when more than 2 dealers are working on the globe, namely, continuous tick data can be obtained except for weekends. The tick data monitor several representative dealers' transaction and the number of data is about 12,000 every day. We analyzed a half year data from October '98 to March '99, and the total number of ticks exceeds a million.

Fig.1 shows a typical example of yen-dollar rate changes in 3 different time scales. Intuitively this figure demonstrates a fractal property of exchange rates in the time axis measured by ticks, namely, Mandelbrot's classical finding also holds for this contemporary market price fluctuation. The statistics of this fluctuation is very close to random walk, actually it is easy to confirm that the power spectrum of this fluctuation clearly follows an inverse square law, that is almost identical to a Brownian motion. The corresponding auto-correlation function for rate fluctuation per tick defined by the following equation decay very quickly as shown in Fig.2.

$$C(T) = \frac{<\Delta r(T_0 + T)\Delta r(T_0)> - <\Delta r(T_0)>^2}{<\Delta r(T_0)^2> - <\Delta r(T_0)>^2} \qquad (1)$$

where $\Delta r(T)$ denotes the rate change at $T$-th tick, namely, $\Delta r(T) = r(T) - r(T-1)$ with $r(T)$ being the exchange rate at $T$-th tick time, and $<\ldots>$ shows average over tick times $T_0$. The correlation is virtually 0 even at $T=2$, which corresponds to the averaged physical time of only 15 seconds. As known from this property the rate change fluctuation is very close to a white noise.

The distribution of rate change in a tick is shown in Fig.3 in semi-log plot with a best-fitted normal distribution for comparison. The tails are much fatter than the normal distribution and we can find extreme events deviating from the mean value, $<\Delta r> = 0.0$, with magnitude nearly 50 times larger than the standard deviation. In order to clarify the functional form of this distribution we plot a cumulative distribution in log-log scale in Fig.4. The distribution for $\Delta r$ in the negative part is plotted by taking absolute values. The distribution functions for both tails are quite identical except for the very large events for which the number of data is limited. The tails decay gradually in log-log scale and we can not find a linear slope. We can approximate the whole shape of the distribution by a symmetric stretched exponential function as

$$P(\geq |\Delta r|) = 0.28 \exp[-7.0(|\Delta r| - 0.04)^{0.4}] \quad , \quad |\Delta r| \geq 0.05 \tag{2}$$

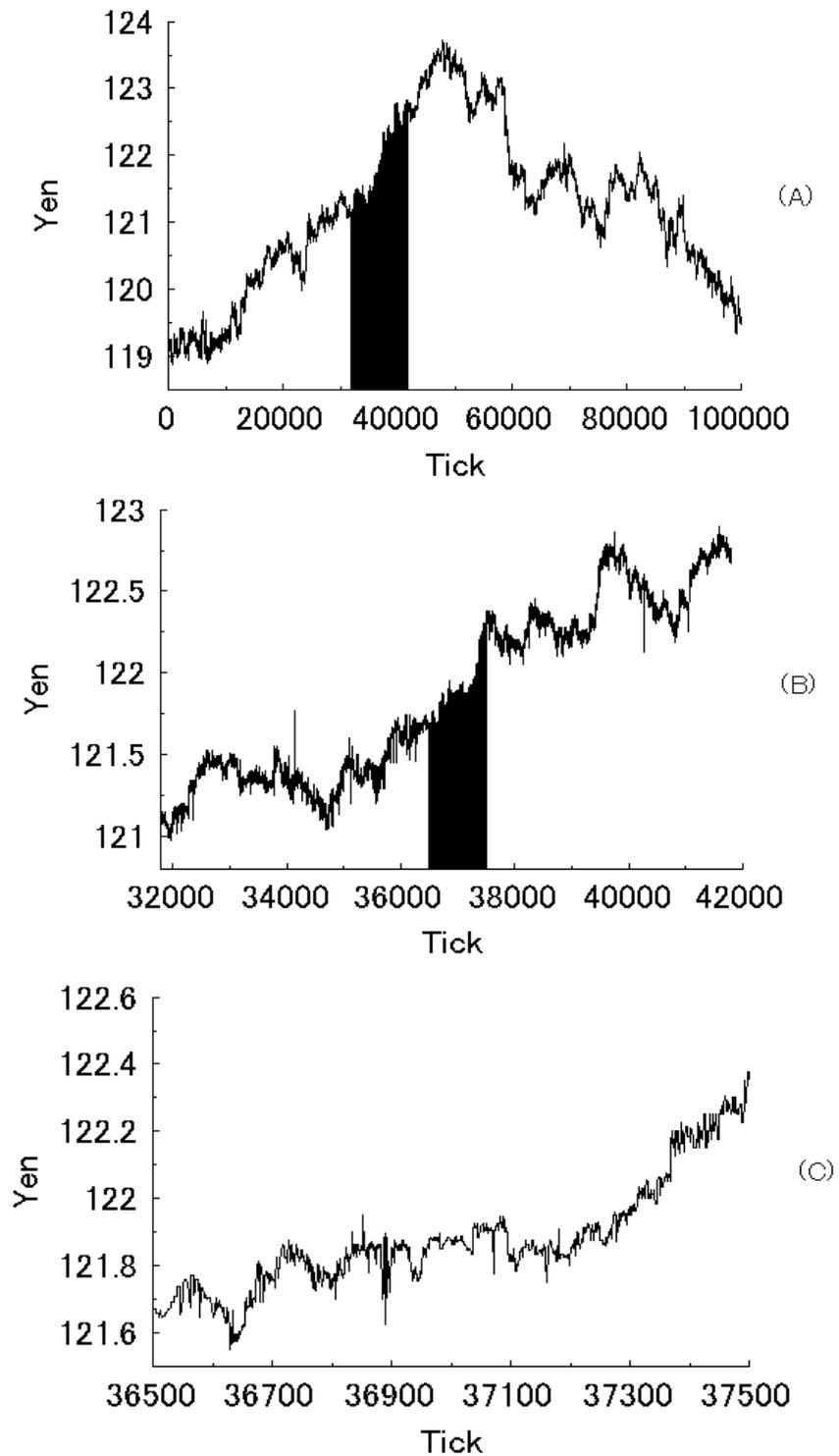

Figure. 1 Fractal property in foreign currency exchange rate fluctuations.
The dark part of (a) is magnified 10 times in (b), and the dark part in (b) is magnified in (c).

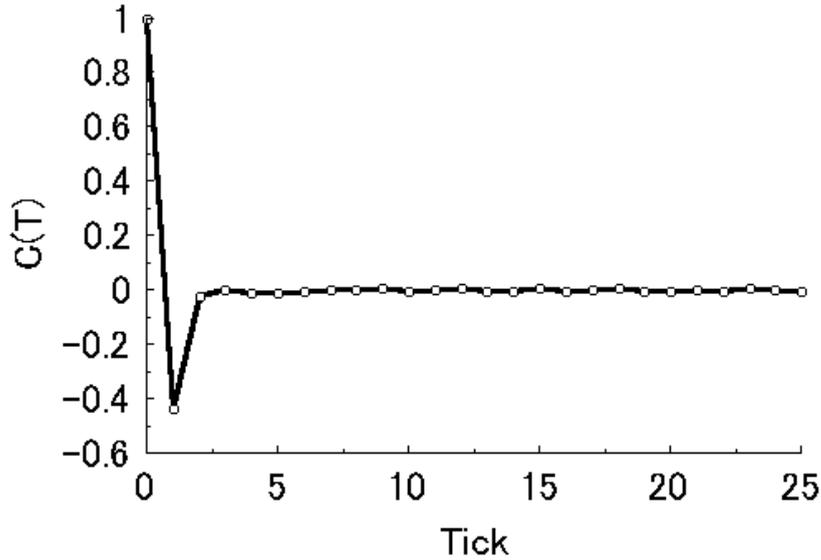

Figure. 2 Autocorrelation of rate changes observed by tick data.

The dashed curve gives this function which is plotted with a shift for better comparison. For small absolute values the function does not fit with this analytic form because real data take discrete values which are given by 0.01 times integers. Omitting about 10 points around the center the theoretical fitting is quite plausible in a wide range. Note that this function decays gradually in the log-log plot, so if we dare to fit a linear slope then we might deduce a non-universal power law that might depend on the fitting range.

    We observe daily change of the distributions of rate fluctuations per tick and find that the distributions are always quite symmetric and the functional forms are always similar to Eq.(2), however, the standard deviation can change more than ten times from 0.03 upto 0.36. This phenomenon is related to the problem of long time correlation in volatility fluctuations observed in any open market price [3]. We can easily confirm that volatility measured by the absolute values of the rate changes has a long correlation, however, the details of its functional forms are yet to be clarified.

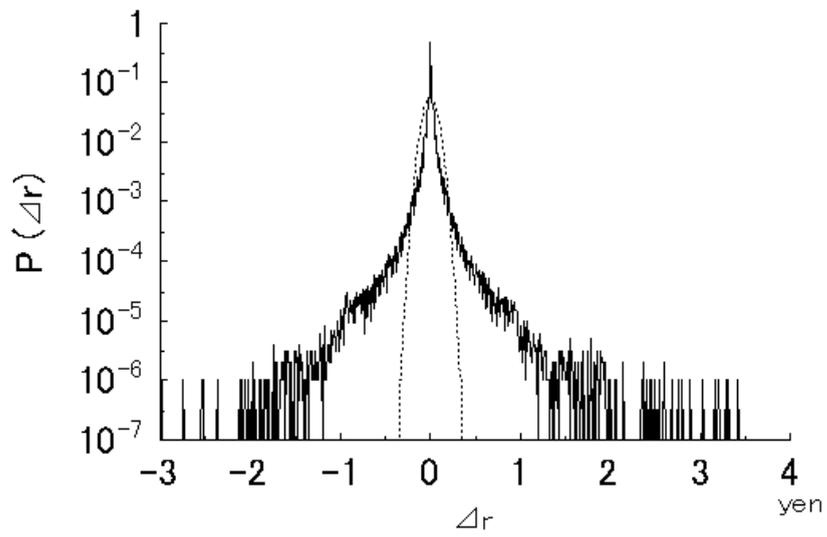

Figure. 3 Rate change distribution viewed by tick data in semi-log plot.

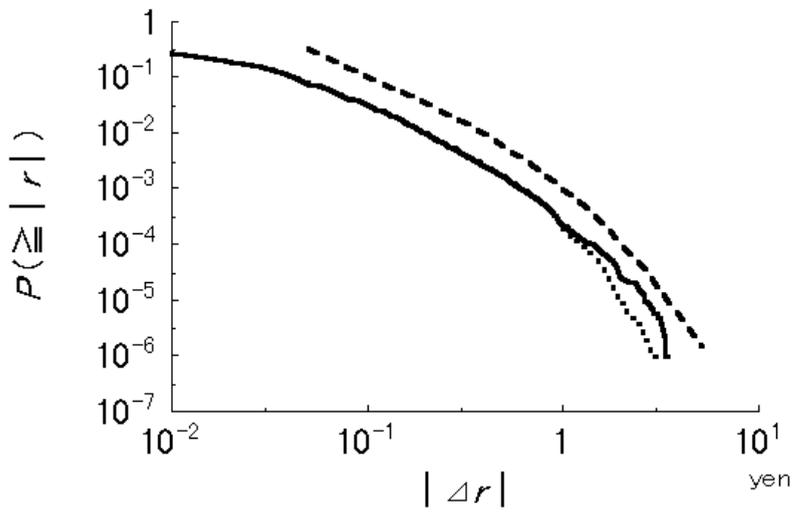

Figure. 4 Log-log plot of the cumulative distribution of rate changes. Positive tail (the bold line), negative tail (dotted line) and the theoretical curve fitted by a stretched exponential function, Eq.(2) (dashed line).

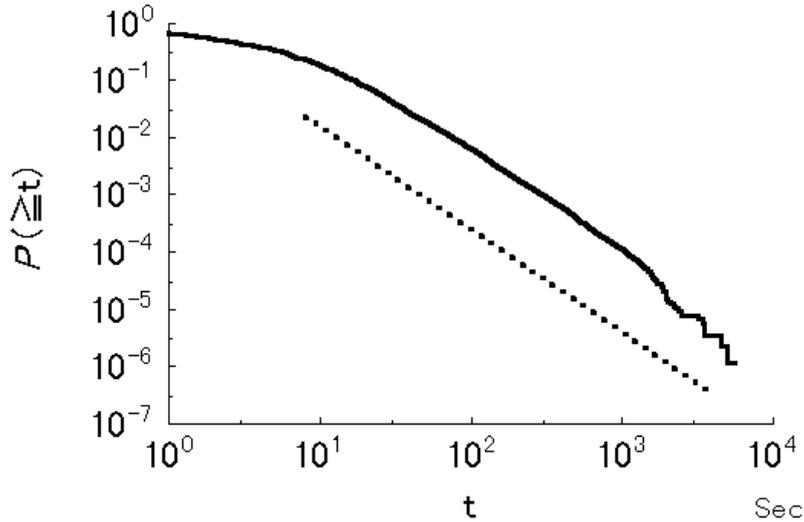

Figure. 5 Distribution of time interval between two successive trades in log-log scale.
The dotted line shows a power law with exponent −1.8.

The time measured by tick is obviously different from the physical time. The occurrence of trades is also a stochastic process and we observe the distribution of tick intervals. In Fig.5 the cumulative distribution of tick intervals measured by seconds are plotted in log-log scale. Here, the minimal tick interval is 0 second and the largest is about 5600 seconds, and intervals longer than this are neglected because they include weekends or holidays. The scale range is therefore limited, however, we can find a power law tail in the scale range of [30, 1600] in seconds.

These statistical properties are not limited to yen-dollar rates but they are applicable to other major currency exchanges. Similarities can be extended to stock markets or other open markets to some extent, however, in such markets trading volumes are much smaller than currency exchange. The total money flow in foreign currency market in one day is estimated to be more than $2*10^{14}$ yen which is roughly a half of the annual gross domestic product (GDP) of Japan. Other than foreign exchanges the markets are open only in office hours, so the statistics are expected to be less universal than the case of currency markets. For example, currency market fluctuations are generally symmetric, however, in stock markets there sometimes appear asymmetric properties such as crashes. We believe that universal properties can be more clearly observed in the foreign currency exchange markets than any other markets.

The fat tail properties in the rate changes can be explained by introducing a simple mathematical model. It is deduced from the analysis of numerical market model that the market price changes can be approximately described by a Langevin equation with fluctuating coefficient [9].

$$\Delta r(t+\Delta t) = b(t)\Delta r(t) + f(t) \qquad (3)$$

Here, in this formulation the time $t$ can be either tick time or physical time, $f(t)$ and $b(t)$ represent the random force term and random coefficient, respectively. The random force term comes from the scatter of dealers' characters and predictions. The value of $b(t)$ is larger when the dealer's averaged responses to the latest price change is larger. It is known that statistical properties of $\Delta r(t)$ are fully characterized by the statistics of $b(t)$ [10]. In the case that $b(t)$ is always smaller than 1 the distribution of $\Delta r(t)$ is known to follow a stretched exponential form. If the probability of $b(t)$ taking a value larger than 1 is not zero, the distribution of $\Delta r(t)$ has power law tails. The exponent of the power law, $\boldsymbol{b}$, is given by a simple formula,

$$P(\geq |\Delta r|) \propto |\Delta r|^{-\boldsymbol{b}} \quad \text{with} \quad <b(t)^{\boldsymbol{b}}> = 1 \qquad (4)$$

here $<...>$ denotes ensemble average. Eq.(4) is mathematically rigorous in the range $0 < \boldsymbol{b} < 2$. When large values appear successively in $b(t)$, $\Delta r$ is enhanced successively by the recursion of Eq.(3) resulting a large value in $\Delta r$.

An interesting point in this discussion is that Langevin equation is very popular in a wide area of physical systems, therefore, a situation that is similar to the case of open market can be realized in purely physical systems. For example, one of the authors (H.T.) and his collaborators are now developing an electrical circuit that has a randomly fluctuating resistance. Note that the case $b(t)>1$ corresponds to a negative resistance, that is, an amplification, in the continuum limit of Eq.(3). According to our preliminary results the statistics of its voltage fluctuations caused by thermal noise shows similarity with that of price changes or rate changes in open markets [11].

The model equation, Eq.(3), captures a basic aspect of price changes in open market, however, it is obviously too simple. Two of the authors (H.T. and M.T.) have already derived a more general basic equation of price changes by a theoretical consideration of the dynamics of dealers [5]. Skipping the derivation the general equation is a set of linear equation with random force and random coefficients as

$$r(t+\Delta t) = r(t) + a(t)(r^*(t) - r(t)) \qquad (5.a)$$
$$r^*(t+\Delta t) = r^*(t) + f(t) + b(t)(r(t) - r(t-\Delta t)) \qquad (5.b)$$

where $r^*(t)$ is a virtual equilibrium price that is determined by the balanced price of demand and supply when all dealers show their buying and selling prices in mind openly. The coefficient $a(t)$ shows the market price's response against the change of demand and supply that is inversely proportional to the price rigidity in economics, namely, for larger $a(t)$ the market price changes more largely for the same modification of demand-supply configuration. For given $\{a(t)\}$ and $\{b(t)\}$ with initial condition for $r(t)$ and $r^*(t)$, Eq.(5a) and (5b) solve the time evolution of $r(t)$ and $r^*(t)$ simultaneously. In the special limit that $r(t)$ is always nearly equal to $r^*(t)$ these set of equations become identical to Eq.(3), then the basic properties of Eq.(3) are recovered.

There are 3 typical behaviors in the time evolution of Eqs.(5). One is a stable state in which the deviation between $r(t)$ and $r^*(t)$ does not increase on average which is realized when both $a(t)$ and $b(t)$ are smaller than 1. Another behavior is an exponential growth of $r(t)$ which corresponds to the phenomenon called "bubble". This behavior occurs when $b(t)$ is larger than 1 and $a(t)$ is smaller than 1, namely when the dealers are strongly expecting the latest trend to be continued and the price rigidity is rather high. In such case the expected equilibrium price *r\*(t)* goes ahead while *r(t)* follows it, and the deviation between *r(t)* and *r\*(t)* increases exponentially. The third behavior is an oscillatory fluctuation that happens when *a(t)* is larger than 1. In such case the price rigidity is so small that the market price goes over *r\*(t)* causing an oscillation.

There are two obvious defects in the simplest model equation, Eq.(3); the lack of long time correlation in volatility and the absence of characteristic behaviors such as bubbles and oscillations which are observable in real open market data. We are now studying the basic properties of the generalized version, Eqs.(5), comparing with real data, so that the values of the coefficients, *a(t)* and *b(t)*, can be estimated from the real time sequences of *r(t)*.

## 4   Company's macroscopic financial data analysis

History of study on fractal properties in money flow can be traced back more than 100 years. In 1896 Pareto investigated personal incomes in Italy and found that the distributions are approximated by a power law [12]. In 1922 Gini explored the same quantity for several European countries and reported that the exponents of the power laws are different for different country [13]. Two physisists, Montroll and Schlesinger, paid attention to the power law nature of income distribution in 1983 and found that personal income of top 1% follows a power law while the rest of 99%, who are mostly salaried, are characterized by a log-normal law [14]. Recently, fractal properties of company's income and asset are attracting much interest.

M. H. R. Stanley et al established an interesting scaling relation on the statistics of growth of companies [15]. Let *A(t)* be the asset of a company at the *t*-th year, where asset means intuitively the price of the company as a whole. Asset growth is characterized by the growth rate that is defined by logarithm of the ratio of asset change,

$$R(t) = \log \frac{A(t)}{A(t-1)} \qquad (6)$$

The distribution function of *R(t)* is observed for more than 10,000 companies in USA and for about ten years. It is shown that the probability density of *R(t)* for a company with asset *A* is approximately given by an exponential function

$$p(R \mid A) = \frac{1}{\sqrt{2}\boldsymbol{s}(A)} \exp(-\frac{\sqrt{2} \mid R - <R(A)> \mid}{\boldsymbol{s}(A)}) \qquad (7)$$

where $\boldsymbol{s}(A)$ is the standard deviation for asset size around *A(t)=A*. There exists a non-trivial scaling relation in $\boldsymbol{s}(A)$ such as

$$\boldsymbol{s}(A) \propto A^{-\boldsymbol{q}} \quad , \quad \boldsymbol{q} \sim 0.15 \qquad (8)$$

Namely, for larger asset companies the fluctuations are relatively smaller, which is intuitively very natural.

There are several quantities that characterize the size of companies such as the number of employee and net sales, and it is confirmed that characterization by using these quantities also leads the same result. Also this type of scaling relation is known to hold not only in USA but also for other countries, for example, Japan, France and Italy [16].

If all of the components of a company fluctuate independently then it is easy to show that the value of the exponent *q* should be 0.5, while if the growth is merely proportional to the whole asset then *q* should be 0. Therefore, the non-trivial value of *q* strongly suggests that the company growth is governed by non-trivial interactions either internal origins [17] or external ones [16].

Universality of the scaling relation, Eq.(8), is still wider. It is known to be valid also for the growth of gross domestic product (GDP) of more than 150 countries for the periods of more than 40 years after the world war II [18]. Although the detail mechanisms of these phenomena are yet to be clarified it is likely that gigantic economic systems seem to fall in the same universality class in which non-trivial fractal relation holds.

Very recently two of the authors (H.T. and M.T.) together with Okuyama examined detailed data of annual incomes of companies in Japan which had been reported to the tax office and they found that a clear power law with exponent very close to -1 holds (see Fig.6) [19]. The data cover all Japanese companies whose incomes are larger than $4.0*10^7$ yen, and the number of the companies is about 85,000. Here, income is in rough sense given by the total sales subtracted by the total outgoing costs. There are fluctuations for very large incomes, however, for companies having income less than $10^{10}$ yen the plots are clearly on a straight line in the log-log plot. Such power law behaviors can also be confirmed for other years and for several other countries.

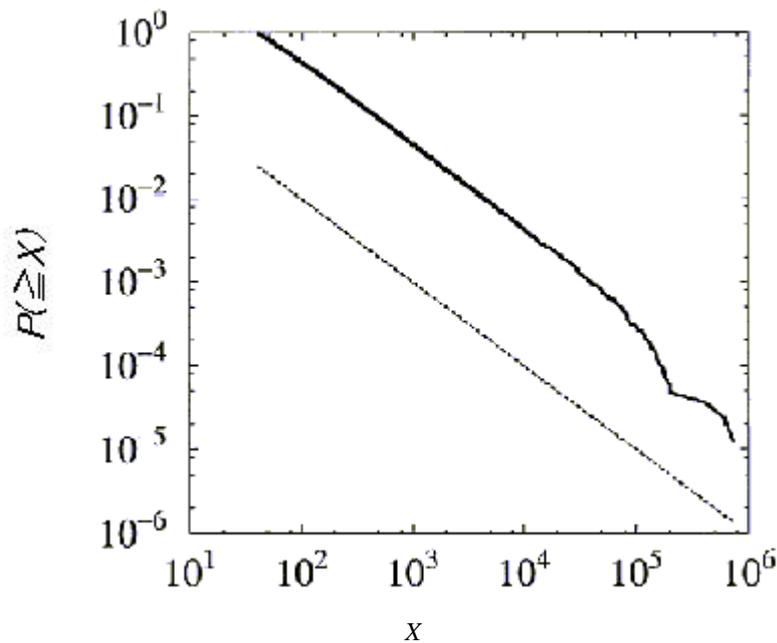

Figure. 6 Log-log plot of the income distribution of Japanese companies for all job categories.
Income *x* is represented in the unit of million yen.

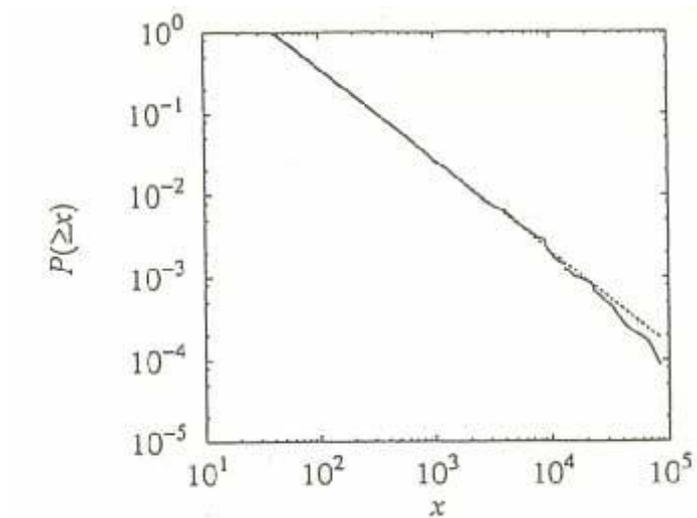

Figure. 7 Income distribution for the category "constructions".

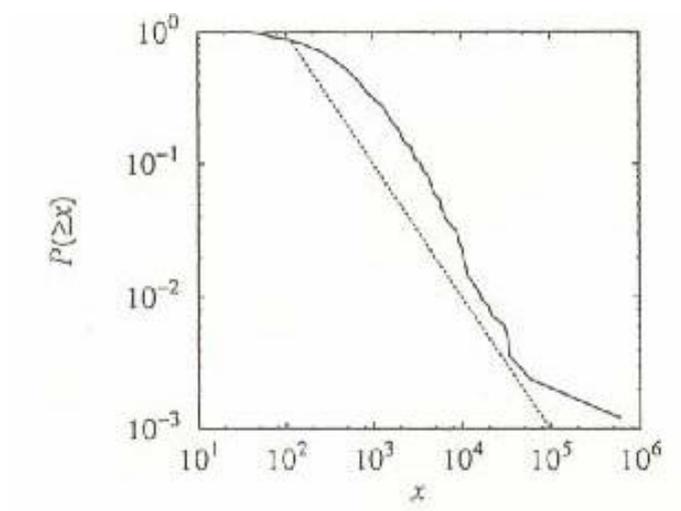

Figure. 8 Income distribution for the category "banks".

An interesting feature of this power law distribution is that similar power law holds in each category of industry. For example, Fig.7 shows the same plot as in Fig.6 but the companies are limited only those categorized in "constructions". The number of companies is about 11,000 and the estimated power exponent is -1.1, very close to the case of the whole categories.

By carefully observing the distributions of income in each job category we find that there are some exceptions that do not follow the power law. For example, in Fig.8 we show the distribution of the case of category "banks". Here, we can not find a clear slope in the plot. The existence of bending point around $10^2$ million yen implies that the number of banks whose incomes are less than this amount is very little. In Japan there were many regulations by law for banks for example the interest rates were standardized all over the country, which causes little competition among banks, also a new bank could not be established easily. As symbolized in this example deviation from power law distributions tends to be observed in job categories which are less competitive compared with other job categories. Recently, Japanese bank regulations have been loosened drastically, so the future change in the distribution of income of banks should attract not only economists' but also scientists' attention.

What is the relation between asset and income both are characterizing company's statistics ? To be precise the income defined above includes taxes to be paid to the government that is about 50% independent of the asset or income size in Japan. The rest of the income is normally added to the company's asset, namely, the following relation holds in rough estimation.

$$A(t+1) = A(t) + I(t)/2 \qquad (9)$$

where $I(t)$ denotes the income of the $t$-th year. It is confirmed from the data that $I(t)$ is generally about 2 orders smaller in magnitude and it can take a negative value, while the company collapses when $A(t+1)$ becomes negative. Comparing Eq.(9) with Eq.(6) we have a relation between $R(t)$ and $I(t)$ as

$$R(t) \propto \frac{I(t)}{A(t)} \qquad (10)$$

The relation between income and asset can be compared directly by plotting these two quantities. Scatters are generally large but the following nonlinear relation has been found as average [19].

$$I(t) \propto A(t)^{d} \quad , \quad d = 0.85 \qquad (11)$$

These relations (10) and (11) are consistent with Eq.(8) and we confirm a relation between the exponents, $d = 1 - q$. This nonlinear relation implies that large systems are less efficient, which is intuitively well appreciated.

So far no mathematical model is known that reproduces all of these empirical relations of company's financial data. Here, we introduce a simplest mathematical model based on competitive growth that realizes the power law distribution with exponent -1. Let us consider a 2-dimensional lattice and we regard each site as a source of unit income. Initially all the sites on the top line are occupied by different companies, therefore, incomes of all companies are 1. In the second step, a company on the top line is chosen randomly and it occupies a site on the second line, and repeat this process until all sites on the second line are occupied. In the third step a company is chosen randomly from the companies on the second line with the probability weight proportional to the total number of occupied sites on the top and second lines, and let it occupy a site on the third line randomly. Repeat this procedure until all sites on the third line are occupied. The same

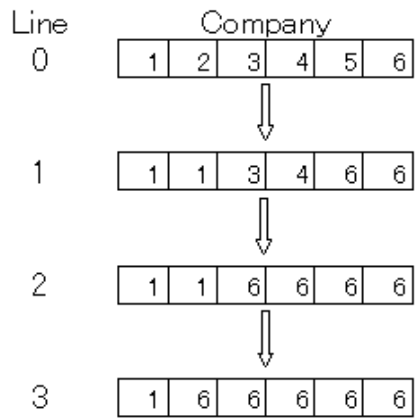

Figure. 9  The company growth model based on competitive growth.

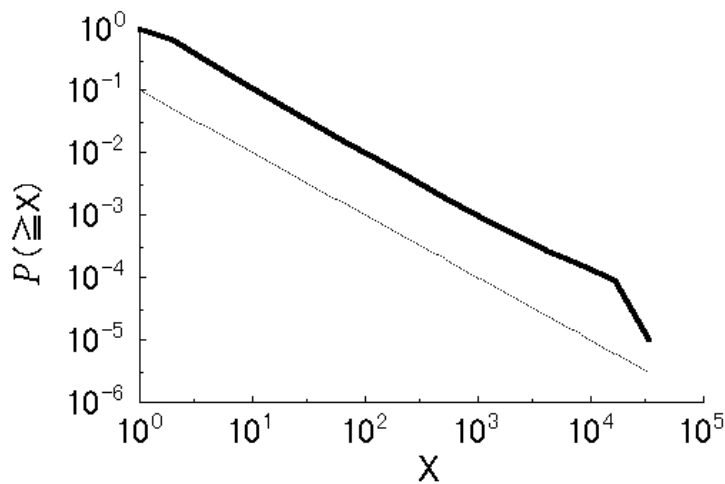

Figurer. 10 Log-log plot of income distribution by the growth model.

growth process is repeated for many lines (see Fig.10). Note that if a company shares no site on the k-th line, then the company stops growing any more.

In this model each new occupation can be regarded as a unit investment that works. This growth process is based on the intuition that a company's income grows with probability proportional to investment and the amount of investment is assumed to be either 0 or proportional to the whole income. The result of this numerical simulation is shown in Fig.9. We can find a clear power law with exponent $-1$, which is consistent with Fig.6. Of course this model captures just one aspect of the phenomenon and we need to develop a model that explains both the asset and income statistics.

## 5 Discussions

We have introduced three topics in this paper; the balance of demand and supply, price fluctuations in open markets and company's macroscopic financial data analysis. These topics may look very different superficially, however, they are deeply interconnected by the key words "fluctuations". Fluctuations had been ignored in economics especially in the discussion of demand and supply, however, as we reviewed in the second chapter the magnitude of fluctuations becomes very large and quick at the balanced point of demand and supply. From the viewpoint of statistical physics the balanced point is the critical point of the underlying phase transition and the fluctuations tend to show fractal characteristics, namely, the averaged amplitude of fluctuations can be infinite in a theoretically ideal situation.

Economic systems tend to be around the critical point in general. Open markets are typical examples. When there are more buyers than sellers, the market price goes up causing decrease in the number of buyers and increase in sellers. Such well-known pulling back mechanism obviously keeps the system around the critical point. The important fact is that this criticality does not stabilize the absolute value of the market price but only stabilizes the statistics of market price fluctuations to follow the critical fluctuations. As the dealers in open markets tend to care about only the relative market price, whether it goes up or down, so the absolute value of the market price is nearly meaningless for the determination of the market price. Therefore, the market prices generally wander almost randomly having statistics characterized by the critical point.

From the viewpoint of fluctuations in demand and supply, company's statistics may also be considered in a similar way. When there appears a new demand in a new field of industry, companies competitively try to supply it. The winner will get a big amount of income and the company will grow. The followers will share the rest of demand and the total supply will increase. When the sum of supply becomes close to the demand, some companies can not gain their incomes and the growth of supply is weakened. As a result demand and supply may nearly balance on average and the whole system may show critical characteristics such as fractal distributions.

The traditional economics theory of demand and supply neglects the effect of fluctuations at all, while the modern financial technology is based on continuous independent Gaussian process. From the standpoint of statistical physics these two cases are extreme limits which correspond to the cases of absolute zero and infinite temperatures, respectively. However, interesting phenomena such as phase transition and pattern formation can never happen in such situations in the case of physical systems. As we have shown in this paper, we can find critical behaviors in real economic data, which considerably resemble physical systems' phase transition phenomenon. It is very likely that there hidden in economic data many more examples which possess fractal properties, because the balance of demand and supply automatically attract any economic system to be around the critical point. Data analysis methods and theoretical formulations developed for phase transition phenomena in physics will become more and more important in economics in the near future.

## 6 Acknowledgements

The authors acknowledge Mr. Kenji Okuyama for his help on preparing some of the figures.

**References**

1. B. B. Mandelbrot, The variation of certain speculative prices, J. of Bussiness (Chicago) **36** (1963) pp.394-419.
2. J. Kertesz and I. Kondor (Eds.), *Econophysics : an emerging science* ( Kluwer Academic Publisher, Dordrecht, to appear).
3. R. Mantegna and E. H. Stanley, *Introduction to Econophysics* (Cambridge Univ. Press, London, to appear).
4. H. Takayasu, A.-h. Sato and M. Takayasu, Power law behaviors of dynamic numerical models of stock market prices, (to appear in ref. 2)
5. H. Takayasu and M. Takayasu, Critical fluctuations of demand and supply, *Physica A* **269** (1999) pp.24-29.
6. D. Sornette, D. Stauffer and H. Takayasu, Maket fluctuations II: multiplicative and percolation models, size effects and predictions, (*Procedings book of conference in Rauischholzhausen 1999*, to appear ( Eds. A. Bunde ) ( http://xxx.lanl.gov/abs/condmat/9909439)).
7. H. Takayasu, H. Miura, T. Hirabayashi and K. Hamada, Statistical properties of deterministic threshold elements – the case of market price, *Physica A* **184** (1992) pp.127-134.
8. R. N. Mantegna and H. E. Stanley, Scaling behavior in the dynamics of an economic index, *Nature* **376** (1995) pp.46-49.
9. A.-h. Sato and H. Takayasu, Dynamic numerical models of stock market price: from microscopic determinism to macroscopic randomness, *Physica A* **250** (1998) pp.231-252.
10. H. Takayasu, A.-h. Sato and M. Takayasu, Stable infinite variance fluctuations in randomly amplified Langevin systems, *Phys. Rev. Lett.* **79** (1997) pp.966-969.
11. A.-h. Sato and H. Takayasu, preparing for publication.
12. V. Pareto, *Le Cour d'Economie Politique*, (Macmillan, London, 1896).
13. C. Gini, Indici di concentrazione e di dipendenza, *Biblioteca delli'economista* **20** (1922).
14. E. W. Montrol and M. F. Shlesinger, *J. Stat. Phys*. **32** (1983) 209.
15. M. H. R. Stanley, L.A.N. Amaral, S. V. Buldyrev, S. Havlin, H. Leschhorn, P. Maass, M. A. Salinger and H. E. Stanley, Scaling behavior in the growth of companies, *Nature* **379** (1996) 804.
16. H. Takayasu and K. Okuyama, Country dependence on company size distributions and a numerical model based on competition and cooperation, *Fractals* **6** (1998) pp.67-79.
17. L. A. N. Amaral, S. V. Buldyrev, S. Havlin, M. A. Salinger and H. E. Stanley, Power law scaling in a system of interacting units with complex internal structure, *Phys. Rev. Lett.*, **80** (1998) pp.1385-1388.
18. Y. L. Lee, L. A. N. Amaral, D. Canning, M. Meyer and H. E. Stanley, Universal features in the growth dynamics of complex organizations, *Phys. Rev. Lett.* **81** (1998) pp.3275-3278.
19. K. Okuyama, M. Takayasu and H. Takayasu, Zipf's law in income distribution of companies, *Physica A* **269** (1999) pp.125-131.